\documentclass[a4paper,fleqn,usenatbib]{mnras}

\usepackage{txfonts}
\usepackage{mathabx}

\usepackage[T1]{fontenc}
\usepackage{ae,aecompl}

\usepackage{booktabs,caption,fixltx2e}
\usepackage[flushleft]{threeparttable}

\usepackage{graphicx}	% Including figure files
\title[Universe opacity and EBL]{Universe opacity and EBL}

\author[V. Vavry\v{c}uk]{
V\'{a}clav Vavry\v{c}uk,$^{1}$\thanks{E-mail: vv@ig.cas.cz}
\\
$^{1}$Institute of Geophysics, The Czech Academy of Sciences, Bo\v{c}n\'{i} II, Praha 4, 14100, Czech Republic\\
}

\date{Accepted 2016 November 1. Received 2016 October 22; in original form 2016 June 28}

\pubyear{2016}

\begin{document}
\label{firstpage}
\pagerange{\pageref{firstpage}--\pageref{lastpage}}
\maketitle

%--------------------------------------------------------------
% Abstract of the paper
%--------------------------------------------------------------
\begin{abstract}

The observed extragalactic background light (EBL) is affected by light attenuation due to absorption of light by galactic and intergalactic dust in the Universe. Even galactic opacity of $10-20$ percent and minute universe intergalactic opacity of $0.01\,\mathrm{mag}\,h\,\mathrm{Gpc}^{-1}$ at the local Universe have a significant impact on the EBL because obscuration of galaxies and density of intergalactic dust increase with redshift as $\left(1+z\right)^3$. Consequently, intergalactic opacity increases and the Universe becomes considerably opaque at $z > 3$. Adopting realistic values for galactic and intergalactic opacity, the estimates of the EBL for the expanding dusty universe are close to observations. The luminosity density evolution fits well measurements. The model reproduces a steep increase of the luminosity density at $z<2$, its maximum at $z=2-3$, and its decrease at higher redshifts. The increase of the luminosity density at low $z$ is not produced by the evolution of the star formation rate but by the fact that the Universe occupied a smaller volume in previous epochs. The decline of the luminosity density at high $z$ originates in the opacity of the Universe. The calculated bolometric EBL ranges from 100 to 200 $\mathrm{n W m}^{-2}\mathrm{sr}^{-1}$ and is within the limits of 40 and 200 $\mathrm{n W m}^{-2}\mathrm{sr}^{-1}$ of current EBL observations. The model predicts 98\% of the EBL coming from radiation of galaxies at $z<3.5$. Accounting for light extinction by intergalactic dust implies that the Universe was probably more opaque than dark for $z>3.5$.  
\end{abstract}

% Select between one and six entries from the list of approved keywords.
% Don't make up new ones.

%--------------------------------------------------------------
\section{Introduction}
%--------------------------------------------------------------

Dust is an important component of the interstellar medium being formed by grains with diameters typically less than 1 $\mu$m. The dust grains interact with the stellar radiation. They absorb and scatter the starlight causing wavelength-dependent light extinction and re-emit the absorbed energy at infrared (IR) and far-infrared (FIR) wavelengths \citep{Draine2003, Draine2011}. The extinction of starlight due to dust and its re-radiation has been observed and modelled by many authors \citep{Mathis1990, Charlot2000, Draine2003, Tuffs2004, Draine2007, Cunha2008, Popescu2011}. Since galaxies contain interstellar dust, they lose their transparency and the starlight of more distant background galaxies is reddened and dimmed when passing through a foreground galaxy \citep{Gonzalez1998, Alton2001}. The reduction of light depends on the galactic opacity, which is controlled by the type of the galaxy, its dust content and the galaxy inclination \citep{Goudfrooij1994, Calzetti2001, Holwerda2005a, Holwerda2005b, Holwerda2007, Lisenfeld2008, Finkelman2008, Finkelman2010}. 

The intergalactic attenuation is much lower than the galactic attenuation and varies with distance from galaxies. High attenuation is observed, for example, in cluster centres being measured by reddening of background objects behind the clusters \citep{Chelouche2007, Bovy2008, Muller2008, Menard2010a} or by evaluating an excess of high-redshift QSOs around low-redshift galaxies  \citep{Boyle1988, Romani1992}. Since the intergalactic absorption of light is very weak, the intergalactic dust is colder than the interstellar dust, and the absorbed energy is re-radiated in the micro-wave spectrum.

The light absorption by galactic and intergalactic dust produces a wavelength dependent galactic and universe opacity, and affects the observed spectral energy distribution as well as the total energy of the extragalactic background light (EBL). The light from distant galaxies might be obscured by foreground galaxies or reduced by intergalactic absorption. Dimming of light by intergalactic absorption is particularly significant for galaxies at high redshifts, because the optical depth of the Universe strongly increases with redshift, see \citet[their Fig. 9]{Menard2010a} or \citet[their Fig. 1]{Imara2016b}. 

This paper is a follow-up to \citet{Vavrycuk2016} where the stellar EBL is studied for the static dusty universe. Here, the luminosity density evolution and the bolometric EBL are studied for the expanding dusty universe. The EBL is calculated for a variety of possible scenarios and the EBL sensitivity is tested to several cosmological parameters. We show how the predicted luminosity density and the EBL fit observations. Finally, we discuss consequences of the universe opacity for the darkness of the early Universe.

%--------------------------------------------------------------
\section{Theory}
%--------------------------------------------------------------

Total (bolometric) energy flux $I$ received per unit area and time from galaxies in an expanding universe is expressed as an integral over redshift $z$ 
\begin{equation}\label{eq1}
I = \int_0^{z_{\mathrm{max}}} \frac{j\left(z\right)}{\left(1+z\right)^2}  
\,e^{-\tau\left(z\right)} \frac{c}{H_0} \frac{dz}{E\left(z\right)} \, \, ,
\end{equation}
where
\begin{equation}\label{eq2}
E\left(z\right) = \sqrt{\left(1+z\right)^2\left(1+\Omega_m z\right)-z\left(2+z\right)\Omega_{\Lambda}}
\end{equation}
is the dimensionless Hubble parameter, $c$ is the speed of light, $H_0$ is the Hubble constant, $j\left(z\right)$ is the luminosity density, $z_{\mathrm{max}}$ is the maximum redshift considered, $\Omega_m$ is the total matter density, $\Omega_{\Lambda}$ is the dimensionless cosmological constant, and $\tau\left(z\right)$ is the redshift-dependent optical depth. Equation (1) is valid for a matter-dominated universe and is identical with the standard formulas (\citeauthor{Peebles1993}\citeyear{Peebles1993}, his Eq. 13.51;  \citeauthor{Dwek1998}\citeyear{Dwek1998}, their Eq. 9; \citeauthor{Peacock1999}\citeyear{Peacock1999}, his Eqs. 3.85 and 3.89) except for the exponential term with optical depth $\tau\left(z\right)$ expressed as 
\begin{equation}\label{eq3}
\tau\left(z\right) = \frac{c}{H_0} \int_0^{z} \left(\frac{\kappa}{\gamma\left(z'\right)} + \lambda\left(z'\right)\right) \,\, \frac{dz'}{E\left(z'\right)} \, \, ,
\end{equation}
where $\kappa$ is the mean opacity of galaxies, $\lambda\left(z\right)$ is the mean intergalactic attenuation along a ray path for galaxies at $z$, and  $\gamma\left(z\right)$ is the mean free path of a light ray between galaxies at $z$
\begin{equation}\label{eq4}
\gamma\left(z\right) = \frac{1}{n \pi a^2} \, \, ,
\end{equation}
where $a$ is the mean galaxy radius, and $n=n\left(z\right)$ is the galaxy number density at $z$. The optical depth (equation 3) comprises the intergalactic light extinction along a ray and the obscuration effect when distant background galaxies are obscured by a foreground galaxy \citep{Harrison1990, Knutsen1997}. The obscuration is weighted by the galactic opacity $\kappa$ which is 1 for a fully opaque galaxy and 0 for a fully transparent galaxy.

If we assume the luminosity of galaxies and the mass within a comoving volume constant in time, expansion of the Universe causes the luminosity density $j\left(z\right)$ in equation (1) to depend on redshift as
\begin{equation}\label{eq5}
j\left(z\right) = j_0 \left(1+z\right)^4  \, \, ,
\end{equation}
where the zero subscript denotes the reference quantity related to the Universe at present. The fourth power of $\left(1+z\right)$  originates in the constant galaxy number density $n$ in the comoving volume causing its apparent increase in the proper volume (the proper number density)
\begin{equation}\label{eq6}
n\left(z\right) = n_0 \left(1+z\right)^3 \, \, ,
\end{equation}
and additionally by an increase of the arrival rate of photons by $\left(1+z\right)$ due to a closer distance between emitting sources at redshift $z$. Equation (5) is well known from observations of the luminosity density at redshifts $z<1$ \citep{Franceschini2001, Franceschini2008, Lagache2005}. However, so far this equation has been interpreted as a consequence of the evolution of the star formation rate, see Section 5.2. Luminosity density evolution.

The difficulties with finding a proper redshift dependence of the luminosity density in the EBL formulas originated in ignoring the increase of the arrival rate of photons by $\left(1+z\right)$ and/or in fixing the reference luminosity density to an early epoch rather than to the present epoch of the Universe. Obviously, fixing to the early cosmic times is possible and mathematically correct \citep[Eq. 3.95]{Peacock1999} but not applicable to calculating the EBL using the luminosity density $j_0$ measured at $z = 0$. 

The change of the proper volume with redshift does not affect the number density $n$ of galaxies only (see equation 6), but also the number density $n_D$ of dust grains, the mean free path $\gamma$, and the intergalactic attenuation $\lambda$ in equation (3), which become the following functions of redshift  
\begin{equation}\label{eq7}
\gamma^{-1} = \gamma_0^{-1} \left(1+z\right)^{3},\,\,\, 
n_{D} = n_{0D} \left(1+z\right)^{3},\,\,\, 
\lambda = \lambda_0 \left(1+z\right)^3 \, \, ,
\end{equation}
where subscript '0' means the quantity at $z = 0$. In addition, the galactic and intergalactic opacities are frequency dependent, according to the '$1/\lambda$ extinction law', where $\lambda$ is the wavelength of light \citep{Mathis1990, Calzetti1994,Charlot2000}. Hence equation (1) is expressed as 
\begin{equation}\label{eq8}
I = \frac{c j_0}{H_0} \int_0^{z_{\mathrm{max}}} \left(1+z\right)^2  
\,e^{-\tau\left(z\right)} \,\, \frac{dz}{E\left(z\right)} \, \, ,
\end{equation}
where effective optical depth $\tau(z)$ reads
\begin{equation}\label{eq9}
\tau\left(z\right) = \frac{c}{H_0} \int_0^{z} \left(\frac{\kappa}{\gamma_0} + \lambda_0\right) \left(1+z'\right)^4 \,\, \frac{dz'}{E\left(z'\right)} \,.
\end{equation}
The term $(1+z')^4$ in equation (9) comprises an increase of the number density of galaxies and of dust grains with $(1+z')^3$ and an increase of galactic and intergalactic opacities with $(1+z')$ due to the $1/\lambda$ extinction law. Since wavelengths measured at $z = 0$ gradually decrease with redshift when going back in time, the opacities increase along a ray.

For transparent galaxies with zero intergalactic attenuation, equation (8) simplifies to
\begin{equation}\label{eq10}
I = \frac{c j_0}{H_0} \int_0^{z_{\mathrm{max}}} \left(1+z\right)^2 \frac{dz}{E\left(z\right)}\,.
\end{equation} 
This integral diverges for infinite $z_{\mathrm{max}}$ for the matter-dominated as well as radiation-dominated universe which looks apparently erroneous and unphysical. The divergence is, however, correct being a consequence of the assumed model which is unphysical. The model predicts an enormously high galaxy luminosity density because of: (1) the time independent mean galaxy luminosity, (2) conservation of the galaxy number density in the comoving volume, and (3) the high concentration of galaxies within a small volume at high redshifts. The galaxy luminosity density is so high that the integrand in equation (10) does not vanish for $z \rightarrow \infty$ and the total EBL summed over all redshifts diverges. 

The divergence of equation (10) disproves the opinion that the Olbers' paradox is eliminated by considering a model of expanding universe of finite age. For example, \citet{Wesson1987} and \citet{Wesson1991} argue that the finite age of the Universe implies that galaxies have not had time to populate the intergalactic space with enough photons to make it bright. The above calculations reveal that such arguments are not correct and that the finite age of the Universe is not a decisive factor. Similarly, the expansion of the Universe does not eliminate the Olbers' paradox, as supposed by some authors \cite[p. 355]{Peacock1999}. The decline of light energy due to the redshift is not enough for suppressing the enormously high light intensity received from high-redshift galaxies.

The divergence of the EBL is removed by considering the obscuration of galaxies and intergalactic attenuation of light due to the partial opacity of the Universe. Since the real values of the galactic opacity and intergalactic attenuation are non-zero, the attenuation-obscuration term in equations (8) and (9) becomes significant at high redshifts and causes the total EBL to be finite. This is clear because the density of the intergalactic matter responsible for the light absorption and the probability that foreground galaxies obscured background galaxies were much higher in the early Universe than at the present epoch \citep[pp 322-323]{Peebles1993}. The Universe is significantly opaque at high redshifts and thus the divergence is eliminated by light extinction. Alternatively, the divergence of the EBL is eliminated if the mean luminosity or the number density of galaxies significantly declines at high redshifts.

%--------------------------------------------------------------
%  Figure 1 (two columns) 
%--------------------------------------------------------------
\begin{figure*}
\centering
\includegraphics[angle=0,width = 16 cm,trim={110 70 140 50},clip]{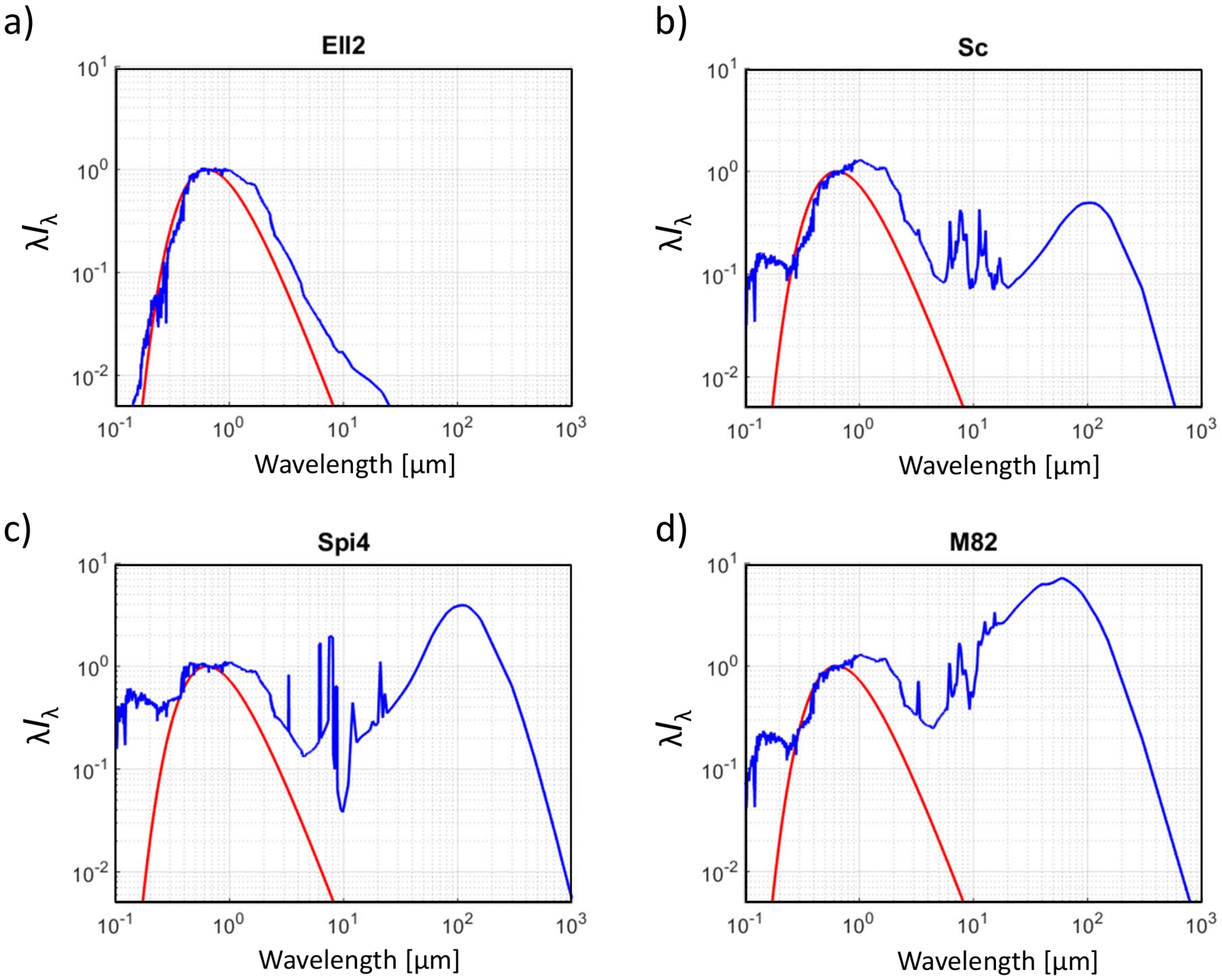}
\caption{
Spectral energy distributions (SED) of four galaxy templates (blue line) of the SWIRE template library \citep{Silva1998, Berta2003} with the SED of Sun (red line). The SEDs are normalized to have the same value at the R band.
}
\label{fig:1}
\end{figure*}

%--------------------------------------------------------------
% Table 1 (2 columns)
%--------------------------------------------------------------
\begin{table*}
%
% table title
%
\caption{Excess ratio for 13 SED galaxy templates of the SWIRE library}  
\label{Table:1}      % is used to refer this table in the text
\centering
\begin{tabular}{c c c c c c c c c c c c c c}  
%
% inserts double horizontal lines
%
\hline\hline                 
%
% table heading 
Galaxy type & Ell2 & Ell5 & Ell13 & S0 & Sa & Sb & Sc & Sd & Sdm & Spi4 & N6090 & N6240 & M82\\
% inserts single horizontal line
%
\hline                        
%
% inserting body of the table
Excess ratio & 1.25 & 1.34 & 1.39 & 1.63 & 1.66 & 1.94 & 2.13 & 2.86 & 3.00 & 5.45 & 8.55 & 10.44 & 10.36 \\     
%
% inserts single horizontal line
%
\hline                                  
\end{tabular}
%
% notes
%
\begin{tablenotes}
\item Extremely high values for star-forming galaxies are not taken into account in further calculations of the mean excess ratio, because their occurrence is statistically insignificant at $z=0$.
\end{tablenotes}
\end{table*}

%--------------------------------------------------------------
\section {Parameters for modelling}
%--------------------------------------------------------------
For calculating the EBL intensity and its evolution in the model of expanding dusty universe we need observations of the mean galactic and intergalactic opacities and the luminosity density and its evolution. 

%--------------------------------------------------------------
\subsection {Galaxy luminosity density and its evolution}
%--------------------------------------------------------------

The luminosity density is a rather well-constrained cosmological parameter standardly determined from the Schechter function  \citep{Schechter1976}. It has been measured by large flux-limited redshift surveys 2dFGRS \citep{Cross2001}, SDSS \citep{Blanton2001, Blanton2003} or CS \citep{Geller1997, Brown2001}. The luminosity function in the R-band was estimated to be $(1.84 \pm 0.04) \times 10^8\, h\, L_{\Sun}\, \mathrm{Mpc}^{-3}$ for the SDSS data \citep{Blanton2003} and $(1.9 \pm 0.6) \times 10^8 \,h\, L_{\Sun}\, \mathrm{Mpc}^{-3}$ for the CS data \citep{Brown2001}.

Since the spectral energy distribution (SED) of galaxies differs from that of stars, the luminosity density $j_0$ for $z=0$ measured at some specified frequency band must further be corrected. The SED of galaxies has remarkably higher values in the infrared spectrum and this 'energy excess' is dependent on the galaxy type (see Fig.~\ref{fig:1}). The correction for the energy excess in the infrared spectrum can be calculated from the bolometric luminosities of a galaxy and Sun $L_G$ and $L_{\Sun}$, normalized to a common R-band value
\begin{equation}\label{eq11}
k_{\mathrm{excess}} = \frac{L_G}{L_{GR}} \, 
\frac{L_{\Sun R}}{L_{\Sun}} \, \, ,
\end{equation}
where $k_{\mathrm{excess}}$ is called the excess ratio. Since the excess ratio is galaxy-type dependent (for basic galaxy templates, see Table~\ref{Table:1}), the mean excess ratio must be calculated by weighted averaging according to relative distribution of individual galaxy types in the Universe. Adopting estimates of the excess ratio from Table~\ref{Table:1}, we get the weighted mean excess ratio in the range of  $k_{\mathrm{excess}} = 1.4 - 2.0$ (see Table~\ref{Table:2}) and the mean bolometric luminosity density at  $z=0$ 
\begin{equation}\label{eq12}
j=  k_{\mathrm{excess}}j_R = 2.5 - 3.8 
 \times 10^8\, h\, L_{\Sun}\, \mathrm{Mpc}^{-3} \,.
 \end{equation}

The luminosity density is not constant but redshift dependent \citep{Madau1998, Kochanek2001, Lonsdale2003}. The observed luminosity density displays a strong increase with redshift, which is best described as $(1 + z)^4$ for $z$ less than 1 \citep{Franceschini2001, Franceschini2008, Hopkins2004, Lagache2005}. This is a luminosity density evolution averaged over different galaxy types. The observed sharply increasing evolution of the luminosity density with redshift means that the background was considerably more powerful in the recent past \citep{Gilmore2012}. The luminosity density culminates at redshifts about 2-3 and then monotonically decreases \citep{Bouwens2007, Bouwens2011}. This decline is observed at all wavelengths \citep{Somerville2012} but the most complete measurements are for the UV luminosity based on the Lyman break galaxy selections. In recent years, the UV measurements were extended for redshifts up to 10-12 \citep{Bouwens2011, Oesch2014, Bouwens2015}. 

%--------------------------------------------------------------
% Table 2 (two columns)
%--------------------------------------------------------------
\begin{table*}
%
% table title
%
\caption{Effective opacity of galaxies and excess ratio}  
\label{Table:2}      % is used to refer this table in the text
\centering
\begin{tabular}{c c c c c c}  
%
% inserts double horizontal lines
%
\hline\hline                 
%
% table heading 
Galaxy type & $w$ & $A_V$ & $\kappa_V$ & $k_{\mathrm{excess}}$ \\  
% units
 & \% & mag &  &  \\     
\hline                        
%
% inserting body of the table
Elliptical & 35 & $0.06 \pm 0.02$ & $0.05 \pm 0.02$ & $1.3 \pm 0.1$ \\ 
Spiral     & 20 & $0.70 \pm 0.20$ & $0.48 \pm 0.15$ & $2.4 \pm 0.7$ \\
Lenticular & 45 & $0.30 \pm 0.10$ & $0.24 \pm 0.08$ & $1.6 \pm 0.3$  \\
Weighted average & &$0.29 \pm 0.09$ & $0.22 \pm 0.08$ & $1.7 \pm 0.3$ \\
%
% inserts single horizontal line
%
\hline                                  
\end{tabular}
%
% notes
%
\begin{tablenotes}
\item $w$ is the frequency of galaxy types in regular clusters, see \citet[Table 4]{Bahcall1999}, 
$A_V$ is the inclination-averaged visual attenuation,
$\kappa_V$ is the visual galactic opacity, and
$k_{\mathrm{excess}}$ is the excess ratio defined in equation (11).
\end{tablenotes}
\end{table*}

%--------------------------------------------------------------
\subsection {Galactic opacity}
%--------------------------------------------------------------

The methods for measuring galactic opacity usually perform multi-wavelength statistical analysis of the colours and number counts of background galaxies produced by a foreground galaxy (for a review, see  \citet{Calzetti2001}). The elliptical galaxies are most transparent with an effective extinction $A_V$ of $0.04 - 0.08$ mag. The spiral and irregular galaxies are more opaque. \citet{Holwerda2005a} reports that attenuation of the disc in the face-on view is formed by two components: the first one is optically thicker ($A_I = 0.5 - 4$ mag) being related to the spiral arms, and the second one is constant and optically thinner related to the disc ($A_I = 0.5$ mag). Typical values for the inclination-averaged extinction are at the B-band \citep{Calzetti2001}: $0.3 - 0.4$ mag for the irregular galaxies, $0.5 - 0.75$ mag for Sa-Sab galaxies, and $0.65 - 0.95$ mag for the Sb-Scd galaxies.

Considering estimates of the distribution of specific galaxy types in the Universe and their mean visual extinctions (see Table~\ref{Table:2}), and recalculating extinctions to opacities 
\begin{equation}\label{eq13}
\kappa_V = 1 - \exp \left(0.9211 A_V \right)\,,
\end{equation}
we can calculate the overall visual opacity of galaxies using weighted averaging  
\begin{equation}\label{eq14}
\left\langle \kappa_V \right\rangle = \sum w_i \kappa_i \, ,
\end{equation}
which is 
\begin{equation}\label{eq15}
\left\langle \kappa_V \right\rangle = 0.22 \pm 0.08 \,.
\end{equation}
%

%--------------------------------------------------------------
\subsection {Universe opacity}
%--------------------------------------------------------------

\citet{Menard2010a} estimated visual intergalactic attenuation to be $A_V = (1.3 \pm 0.1) \times 10^{-2}$ mag at distance from a galaxy up to 170  kpc, and $A_V = (1.3 \pm 0.3) \times 10^{-3}$ mag at distance up to 1.7 Mpc. Similar values are presented by \citet{Muller2008} and \citet{Chelouche2007} for the visual attenuation produced by intracluster dust. However, the intergalactic attenuation increases with redshift. Hence, effectively transparent universe at zero redshift becomes opaque (optically thick) at redshifts of  $z = 1 - 3$ \citep{Davies1997}. The strong increase of intergalactic extinction with redshift is also reported by \citet{Menard2010a} by correlating the brightness of $\sim$85.000 quasars at $z > 1$ with the position of 24 million galaxies at $z \sim 0.3$ derived from the SDSS Survey. The authors obtained extinction $A_V$ of about 0.03 mag at $z = 0.5$ but to about $0.05 - 0.09$ mag at $z = 1$. 

The universe opacity has been intensively studied also in relation to the accelerated expansion of the Universe revealed by unexpected dimming of type Ia supernovae (SNe Ia), see \citet{Riess1998, Perlmutter1999}. The key issue was to estimate a contribution of the universe opacity to SNe Ia dimming. This was done by analysing a violation of the cosmic distance duality (CDD) also known as the Etherington's reciprocity law \citep{Etherington1933}, being valid for all non-dissipative cosmological models based on Riemannian geometry \citep{Ellis2007}. Combining SNe Ia data with measurements of the Hubble expansion, \citet{Avgoustidis2010} confirmed the violation of the CDD and estimated the optical depth of the Universe at visible wavelengths to be ~0.01 at redshifts ranging from 0.2 to 0.35. Similar values have also been obtained by other authors \citep{Nair2012, Holanda2013}. In addition, consistent opacity was recently reported by \citet{Xie2015} who studied the luminosity and redshifts of the quasar continuum at the data sample of ~90.000 objects and estimated the effective dust density $n \sigma_V \sim 0.02 \,h \, \mathrm{Gpc}^{-1}$ at $z < 1.5$. Due to the increase of the dust density with redshift $z$, the extinction magnitude can reach a value of $A_V = 1$ at $z = 3$ \citep[their Fig. 5]{Xie2015}. Since attenuation at infrared wavelengths is much lower than in the visible spectrum, the bolometric attenuation $A$ would be about twice lower than $A_V$ \citep{Mathis1990, Lim2014}.

%--------------------------------------------------------------
\subsection {Opacity ratio}
%--------------------------------------------------------------

Extinction of light of galaxies is caused: (1) by the galactic opacity causing obscuration of background galaxies by partially opaque foreground galaxies, and (2) by the universe opacity produced by light absorption by intergalactic dust. These two effects are responsible for absorption of the EBL energy which is mathematically described by equation (9) for optical depth $\tau\left(z\right)$. The proportion of the EBL energy absorbed by galaxies and by intergalactic dust can be quantified by the so-called opacity ratio

\begin{equation}\label{eq16}
R_\kappa = \frac{\lambda_0 \gamma_0}{\kappa} \,\, ,
\end{equation}
which is a redshift independent constant. If attenuation of light due to galactic and intergalactic dust follows the same extinction law, the opacity ratio becomes also frequency independent.

\begin{figure}
\centering
\includegraphics[angle=0,width = 8.0 cm,trim= {120 80 120 80},clip]{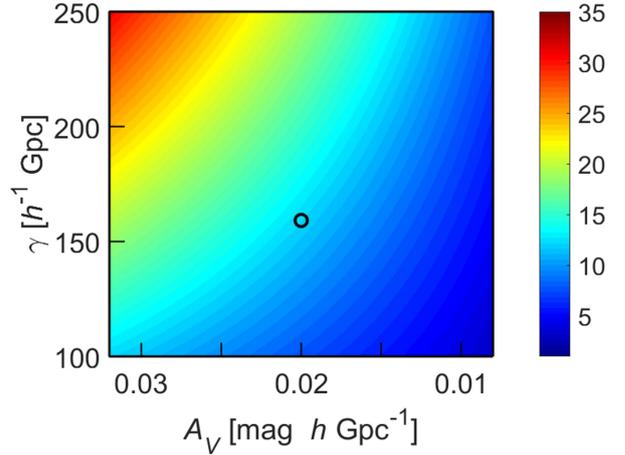}
\caption{
Opacity ratio $R_\kappa$ evaluating the relative impact of the intergalactic opacity and the obscuration of galaxies on the total EBL. The ratio is shown as a function of intergalactic visual attenuation $A_V$ and mean free path between galaxies $\gamma$. The galactic opacity is $\kappa_V = 0.22$. The open circle marks the position for the optimum value $R_\kappa = 13.4$.}
\label{fig:2}
\end{figure}

Considering observations of the galactic and intergalactic opacity, and the mean free path of light between galaxies, the opacity ratio is estimated to be in the range of 5-40 with an optimum value of 13 (see Table~\ref{Table:3} and Fig.~\ref{fig:2}). This indicates that light radiated by galaxies is predominantly absorbed by intergalactic dust. Absorption of light due to galaxy obscuration is much lower.

%--------------------------------------------------------------
%  Figure 2 (two columns) 
%--------------------------------------------------------------

%--------------------------------------------------------------
%  Figure 3 (two columns) 
%--------------------------------------------------------------
\begin{figure*}
\centering
\includegraphics[angle=0,width = 15 cm,trim= {50 80 50 40},clip]{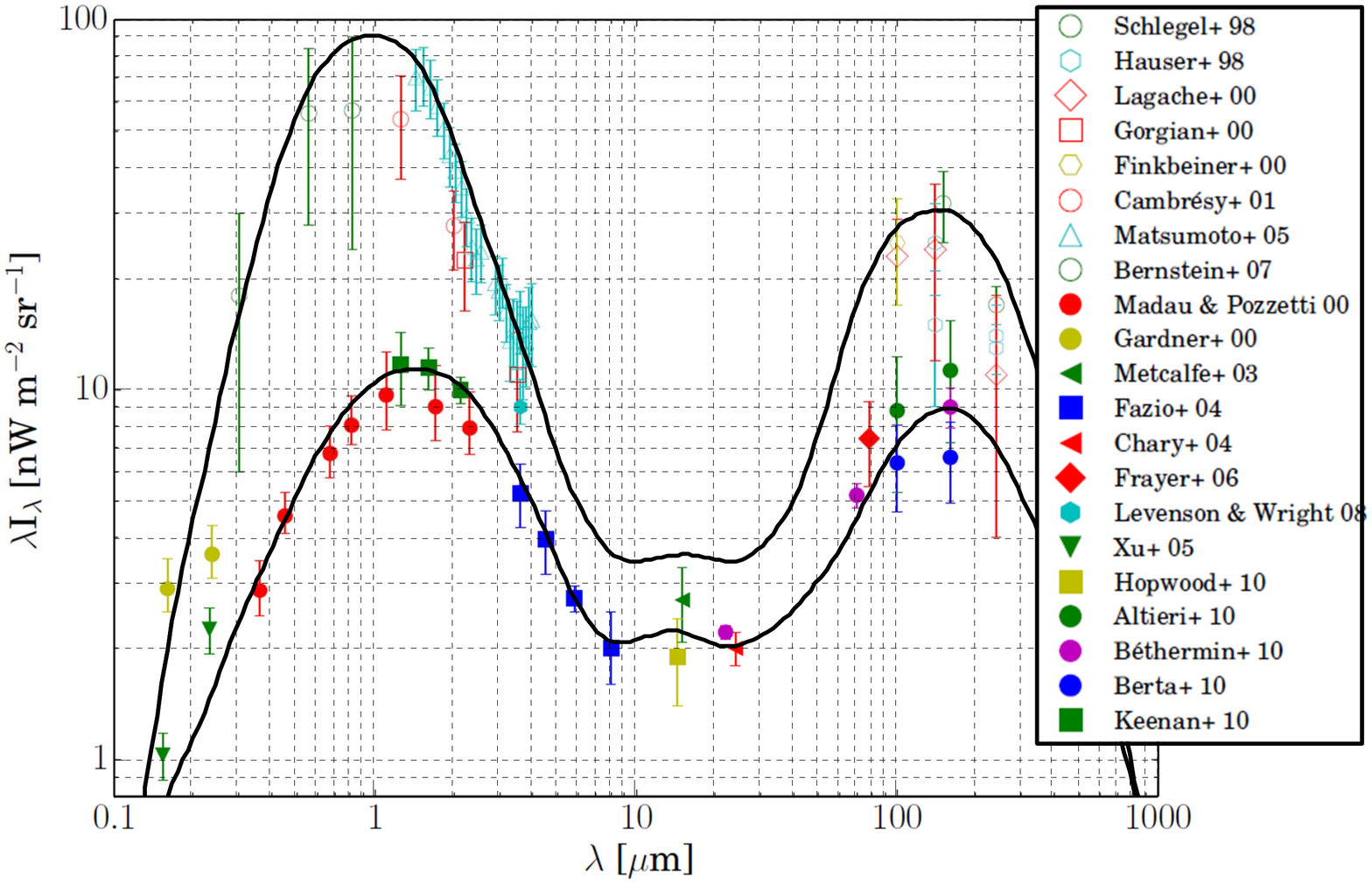}
\caption{
Spectral energy distribution (SED) of the extragalactic background light (EBL) with estimates of its minimum and maximum limits (black lines). The observations reported by various authors are marked by colour symbols (modified after \protect \cite{Dominguez2011}).
}
\label{fig:3}
\end{figure*}
%
%--------------------------------------------------------------
% Table 3 (two columns)
%--------------------------------------------------------------
\begin{table*}
%
% table title
%
\caption{Opacity ratio}  
\label{Table:3}      % is used to refer this table in the text
\centering
\begin{tabular}{c c c c c c c c c}  
%
% inserts double horizontal lines
%
\hline\hline                 
%
% table heading 
Value & $n$ & $\gamma$ & $\kappa_V$ & $A_V$ & $\lambda_V$ & $R_\kappa$  \\  
%
% units
%
 & $h^3\,\mathrm{Mpc}^{-3}$ & $h^{-1}\,\mathrm{Gpc}$ &  & $\mathrm{mag}\,h\, \mathrm{Gpc}^{-1}$ & $ h\, \mathrm{Gpc}^{-1}$  \\
%
% inserts single horizontal line
%
\hline                        
%
% inserting body of the table
Minimum $R_\kappa$ & 0.015 & 130 & 0.30  & 0.015 & 0.0138 & 6.0 \\ 
Maximum $R_\kappa$ & 0.025 & 210 & 0.14  & 0.030 & 0.0276 & 41.4 \\
Optimum $R_\kappa$ & 0.020 & 160 & 0.22  & 0.020 & 0.0184 & 13.4 \\
%
% inserts single horizontal line
%
\hline                                  
\end{tabular}
%
% notes
%
\begin{tablenotes}
\item 
$n$ is the number density of galaxies, 
$\gamma$ is the mean free path between galaxies defined in equation (4), 
$\kappa_V$ is the mean visual opacity of galaxies, 
$A_V$ is the visual intergalactic extinction, 
$\lambda_V$ is the visual intergalactic extinction coefficient, 
$R_\kappa$ is the opacity ratio calculated using equation (16). The mean effective radius of galaxies $a$ is considered to be 10 kpc in equation (4), see \citet{Vavrycuk2016}.
\end{tablenotes}

\end{table*}

%--------------------------------------------------------------
\section {Observations of EBL}
%--------------------------------------------------------------

The EBL covers the near-ultraviolet, visible and near- and far-infrared wavelengths in the range from 0.1 to 1000 $\mu$m. The direct EBL data were provided by the COBE mission, by the ISO instruments at infrared wavelengths, and by the SCUBA instrument at submillimeter wavelengths \citep{Cooray2016}. These measurements are appended by integrating light from extragalactic source counts. The cumulative brightness of galaxies yields a lower limit since the number of unresolved sources is unknown \citep{Madau2000, Hauser2001,Fazio2004, Dole2006, Thompson2007}. The upper limits on the EBL are obtained by analyzing high-energy gamma-rays from distant blazars attenuated by pair production with the EBL photons  \citep{Kneiske2004, Dwek2005, Aharonian2006, Stecker2006, Abdo2010, Primack2011, Gilmore2012}.

The spectral energy distribution of the EBL has two peaks: at visible-to-near-infrared wavelengths (0.7 to 2 $\mu$m) corresponding to stellar light, and at far-infrared wavelengths  (100 to 200 $\mu$m) corresponding to thermal radiation of dust in galaxies  \citep{Schlegel1998, Calzetti2000}. Although the EBL spectrum has been measured by many experiments, the uncertainties are large (see Fig.~\ref{fig:3}). The most significant uncertainties now seem to be in the  $20-100\, \mu$m wavelength range \citep{Bernstein2007}. Integrating the lower and upper limits of the spectral energy distributions shown in Fig.~\ref{fig:3}, the total EBL should fall between 40 and 200 $\mathrm{nWm}^{-2}\mathrm{sr}^{-1}$. The most likely value of the total EBL from 0.1 to 1000 $\mu$m is about 100 $\mathrm{nWm}^{-2}\mathrm{sr}^{-1}$ \citep{Hauser2001, Bernstein2002a, Bernstein2002b, Bernstein2002c, Bernstein2007}.

%--------------------------------------------------------------
\section {Results}
%--------------------------------------------------------------
%--------------------------------------------------------------
% Table 4 (two columns)
%--------------------------------------------------------------
\begin{table*}
%
% table title
%
\caption{Input cosmological parameters and resultant EBL}  
\label{Table:4}      % is used to refer this table in the text
\centering
\begin{tabular}{c c c c c c c c c}  
%
% inserts double horizontal lines
%
\hline\hline                 
%
% table heading 
Value & $A_V$ & $A$ & $j_R$ & $k_{\mathrm{excess}}$ &  $j$ & $z^*$ & $I_{\mathrm{theor}}$ & $I_{\mathrm{obs}}$ \\  
%
% units
%
 & $\mathrm{mag}\,h\,\mathrm{Gpc}^{-1}$ & $\mathrm{mag}\,h\,\mathrm{Gpc}^{-1}$ &   $10^8\,h\,L_{\Sun}\, \mathrm{Mpc}^{-3}$ &  & $10^8\,h\,L_{\Sun}\, \mathrm{Mpc}^{-3}$  & & $\mathrm{nWm}^{-2}\mathrm{sr}^{-1}$ & $\mathrm{nWm}^{-2}\mathrm{sr}^{-1}$\\
%
% inserts single horizontal line
%
\hline                        
%
% inserting body of the table
Minimum EBL & 0.030 & 0.015  & 1.80  & 1.4 & 2.5 & 3.0 & 97 & 40 \\ 
Maximum EBL & 0.015 & 0.0075 & 1.88  & 2.0 & 3.8 & 3.8 & 196 & 200 \\
Optimum EBL & 0.020 & 0.010  & 1.84  & 1.7 & 3.1 & 3.4 & 145 & 100 \\
%
% inserts single horizontal line
%
\hline                                  
\end{tabular}
%
% notes
%
\begin{tablenotes}
\item 
$A_V$ is the visual intergalactic extinction, 
$A$ is the bolometric intergalactic extinction, 
$k_{\mathrm{excess}}$ is the excess ratio defined in equation (11), 
$j_R$ is the R-band luminosity density \citep{Blanton2003},
$j$ is the bolometric luminosity density obtained by multiplying $j_R$ by the excess ratio $k_{\mathrm{excess}}$, 
$z^*$ is the saturation redshift, and
$I_{\mathrm{theor}}$ and $I_{\mathrm{obs}}$ are the predicted and observed EBL intensities. 
All input parameters are taken at the zero redshift, and
$h$ is 67.7.
\end{tablenotes}

\end{table*}

%--------------------------------------------------------------
\subsection {Predicted EBL and saturation redshift}
%--------------------------------------------------------------

%--------------------------------------------------------------
%  Figure 4 (two columns) 
%--------------------------------------------------------------
\begin{figure*}
\centering
\includegraphics[angle=0,width=17 cm,trim = {90 200 70 180},clip]{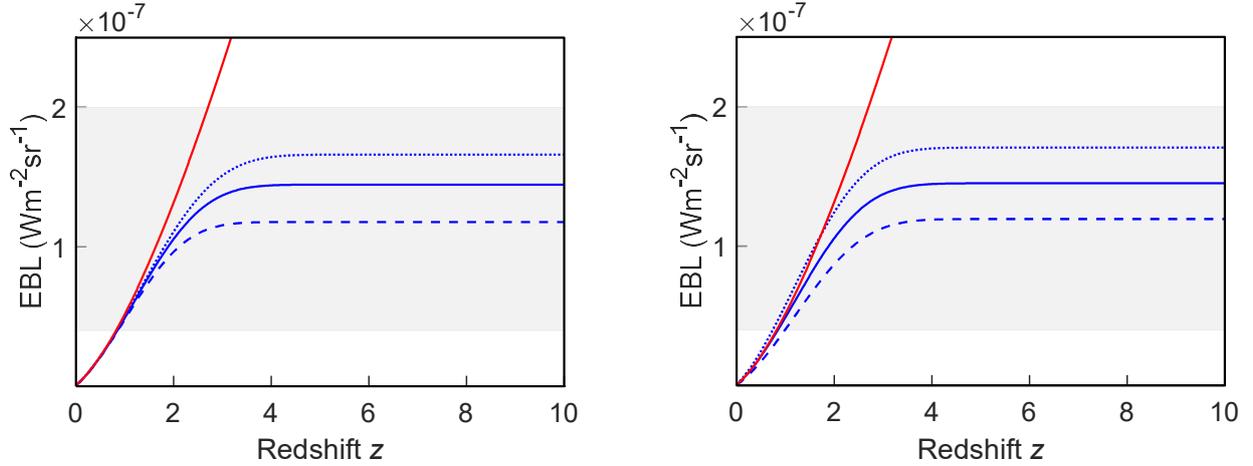}
\caption{
The cumulative bolometric EBL (i.e. the EBL received for galaxies with redshift up to $z$) as a function of redshift $z$. (a) The EBL is shown for the bolometric universe opacity of 0.0075 (blue dotted line), 0.010 (blue solid line) and $0.015 \,\mathrm{mag}\, h\,\mathrm{Gpc}^{-1}$ (blue dashed line). The bolometric luminosity density $j$ is $3.1 \times 10^8 \, h \, L_{\Sun} \, \mathrm{Mpc}^{-3}$. (b) The EBL is shown for the luminosity density $j$ of $2.5 \times 10^8$ (blue dashed line), $3.1 \times 10^8$ (blue solid line) and $3.8 \times 10^8 \, h \, L_{\Sun} \, \mathrm{Mpc}^{-3}$ (blue dotted line). The bolometric universe opacity is $0.01 \, \mathrm{mag} \,h\, \mathrm{Gpc}^{-1}$. For the other parameters, see Table~\ref{Table:4} (optimum values). The red line shows the EBL when the attenuation-obscuration term is neglected (see equation 10). The shadow zone indicates the range of the observed EBL.
}
\label{fig:4}
\end{figure*}

\begin{figure}
\centering
\includegraphics[angle=0,width = 10 cm,trim = {200 150 120 130},clip]{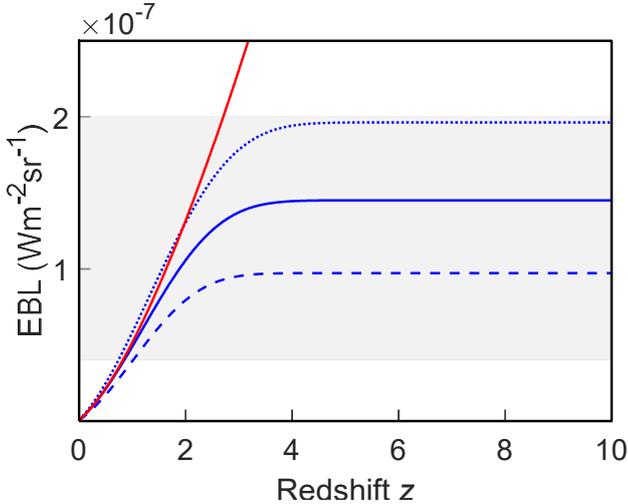}
\caption{
The cumulative bolometric EBL as a function of redshift $z$. The EBL is shown for $A = 0.0075 \, \mathrm{mag}\,h\,\mathrm{Gpc}^{-1}$ and $j = 3.8 \times 10^8 \, h \, L_{\Sun} \, \mathrm{Mpc}^{-3}$ (blue dotted line); $A = 0.010 \,\mathrm{mag}\,h\,\mathrm{Gpc}^{-1}$ and $j = 3.1 \times 10^8 \, h \, L_{\Sun} \, \mathrm{Mpc}^{-3}$ (blue solid line), and $A = 0.015 \,\mathrm{mag}\,h\,\mathrm{Gpc}^{-1}$ and $j = 2.5 \times 10^8 \, h \, L_{\Sun} \, \mathrm{Mpc}^{-3}$ (blue dashed line).
For the other parameters, see Table~\ref{Table:4} (optimum values). The red line shows the EBL when the attenuation-obscuration term is neglected (see equation 10). The shadow zone indicates the range of the observed EBL.
}
\label{fig:5}
\end{figure}

Estimating individual cosmological parameters and their uncertainties from observations (see Table~\ref{Table:4}) we can calculate the EBL intensity using equations (8) and (9) for varying redshift $z_{\mathrm{max}}$. The calculated EBL intensity is mostly sensitive to the universe opacity and luminosity density, see Fig.~\ref{fig:4}. We consider an optimum luminosity density with three values of the universe opacity (Fig.~\ref{fig:4}a) and an optimum universe opacity with three values of the luminosity density (Fig.~\ref{fig:4}b). Fig.~\ref{fig:5} displays the cumulative EBL if both the universe opacity and luminosity density are varying. Figs.~\ref{fig:4} and ~\ref{fig:5} indicate that (1) neglecting the attenuation-obscuration (red lines) produces remarkable effects on the EBL for redshifts $z > 1$, and (2) the EBL is formed mostly by light of galaxies at low redshifts. The contribution of high-redshift galaxies to the EBL is almost negligible. If we define the so-called 'saturation' redshift  $z^*$ as the redshift up to which 98\% of the EBL is received, we get the EBL to be saturated at redshift of   $z^* = 3.4$. If the attenuation-obscuration is neglected, the EBL diverges. 

\begin{figure*}
\centering
\includegraphics[angle=0,width = 17 cm,trim = {40 180 50 120},clip]{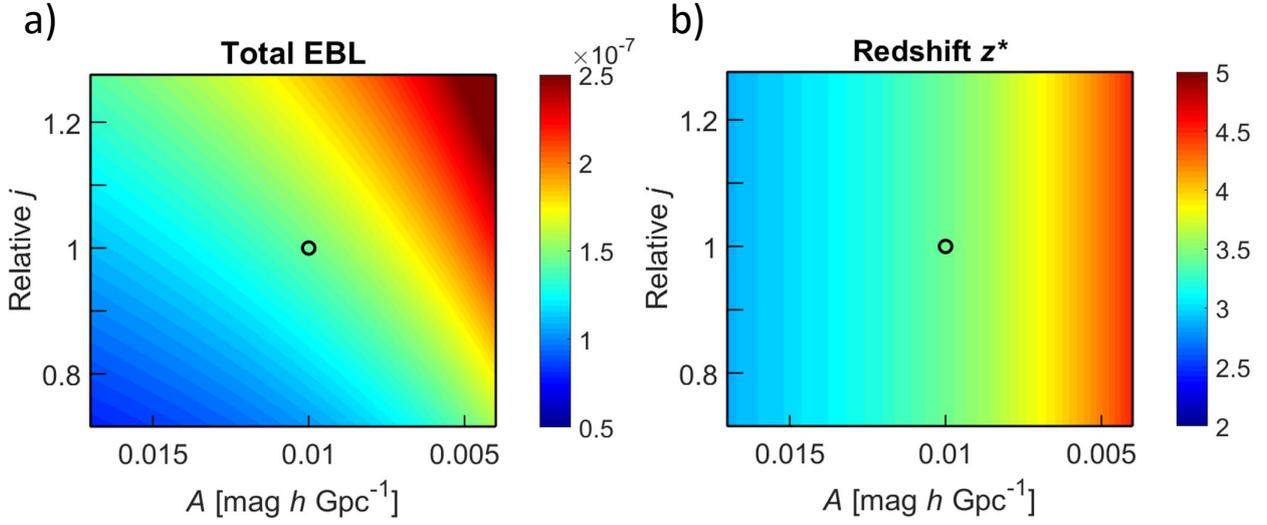}
\caption{
The bolometric EBL (a) and saturation redshift $z^*$ (b) as a function of the universe opacity and the relative luminosity density. The redshift $z^*$ is defined as the redshift at which the cumulative EBL reaches 98\% of its final value. The relative luminosity density is normalized to its optimum value $3.1 \times 10^8 \, h \, L_{\Sun} \, \mathrm{Mpc}^{-3}$ (see Table~\ref{Table:4}). The EBL is in $\mathrm{W m}^{-2}\mathrm{sr}^{-1}$. The black open circles mark the positions of the optimally chosen parameters.
}
\label{fig:6}
\end{figure*}

Figs.~\ref{fig:6}ab show the EBL and saturation redshift $z^*$ for various combinations of the bolometric intergalactic attenuation and the relative luminosity density (i.e. the ratio between the actual and optimum luminosity densities). The EBL varies from 70 to 250 $\mathrm{nWm}^{-2}\mathrm{sr}^{-1}$ with the optimum value of 145 $\mathrm{nWm}^{-2}\mathrm{sr}^{-1}$ (see Fig.~\ref{fig:6}a). The saturation redshift $z^*$ depends mainly on the intergalactic attenuation (see Fig.~\ref{fig:6}b); it varies from 3 to 4.5 with the optimum value of 3.4. It means that the Universe becomes effectively opaque for redshifts higher than $z^* > 3.4$, and the contribution of light of galaxies to the EBL is negligible for galaxies at larger distances.

The predicted bolometric EBL varies roughly from 100 to 200 $\mathrm{nWm}^{-2}\mathrm{sr}^{-1}$ with the optimum value of 145 $\mathrm{nWm}^{-2}\mathrm{sr}^{-1}$ (see Table~\ref{Table:4}). The range of the predicted values of the EBL is broad being produced by a limited accuracy of input parameters. However, the limits for the observed EBL are also large (see Fig.~\ref{fig:3}). The lower and upper limits of the EBL obtained from observations are 40 and 200 $\mathrm{nWm}^{-2}\mathrm{sr}^{-1}$ \citep{Hauser2001, Bernstein2002a, Bernstein2002b, Bernstein2002c, Bernstein2007}, hence the predicted EBL is within the range of current observations. Taking into account all simplifications made in calculations, the agreement is excellent.

%--------------------------------------------------------------
%  Figure 5 (one column) 
%--------------------------------------------------------------
%
%--------------------------------------------------------------
%  Figure 6 (two columns) 
%--------------------------------------------------------------

%--------------------------------------------------------------
%  Figure 7 (two columns) 
%--------------------------------------------------------------
\begin{figure*}
\centering
\includegraphics[angle=0, width = 15 cm,trim = {130 70 70 80},clip]{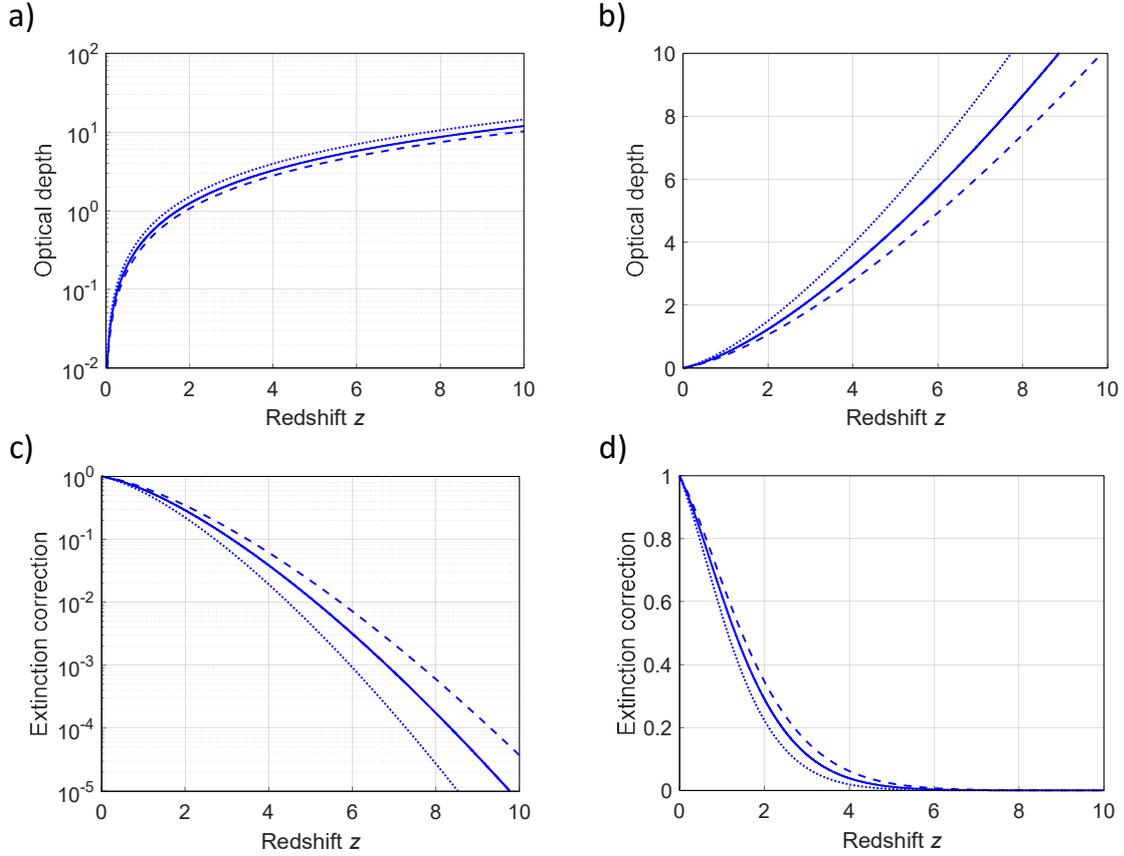}
\caption{
The optical depth (a-b) and the extinction correction (c-d) as a function of redshift $z$. Left plots - the logarithmic scale, right plots - the linear scale. The universe opacity at UV wavelengths $A_{UV}$ is $8.5 \times 10^{-2}$ (blue dotted line), $7.0 \times 10^{-2}$ (blue solid line) and $6.0 \times 10^{-2} \, \mathrm{mag} \, h\, \mathrm{Gpc}^{-1}$ (blue dashed line).
}
\label{fig:7}
\end{figure*}

%--------------------------------------------------------------
%  Figure 8 (one column) 
%--------------------------------------------------------------
\begin{figure}
\centering
\includegraphics[angle=0,width = 9 cm,trim = {140 130 120 130},clip]{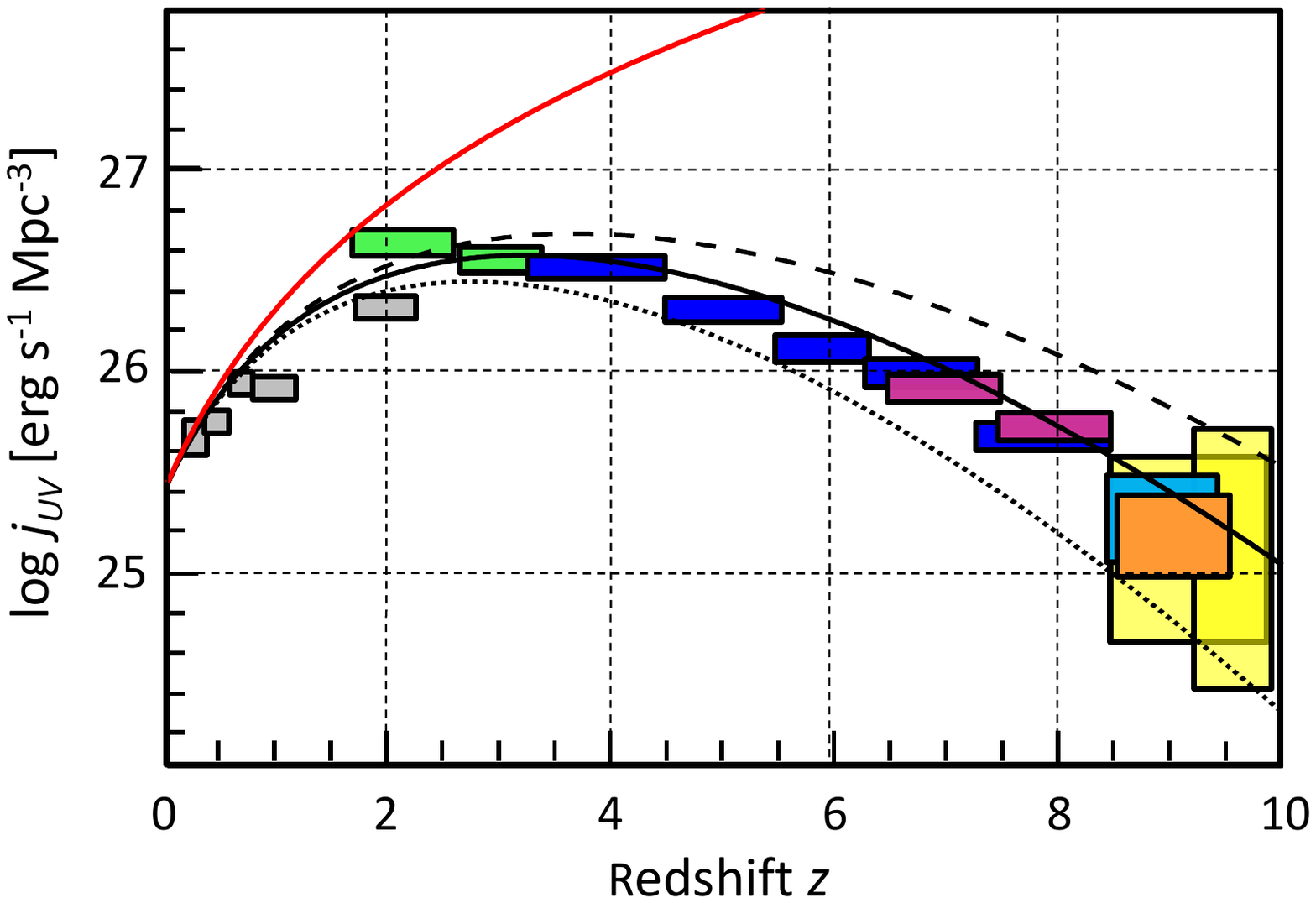}
\caption{
The UV luminosity density as a function of redshift. Observations are taken from \protect \cite{Schiminovich2005} (grey rectangles), 
\protect \cite{Reddy2009} (green rectangles), \protect \cite{Bouwens2014a} (blue rectangles), \protect \cite{McLure2013} (magenta rectangles), \protect \cite{Ellis2013} (orange rectangle), \protect \cite{Oesch2014} (light blue rectangle), and \protect \cite{Bouwens2014b} (yellow rectangles). The predicted luminosity is shown for the transparent universe (red solid line) and the opaque universe with UV intergalactic extinction of 0.06 (dashed black line), 0.07 (solid black line), and 0.085 mag $h\,\mathrm{Gpc}^{-1}$ (dotted black line). The galaxy number density in the comoving volume and the mean galaxy luminosity are assumed to be independent of redshift.
}
\label{fig:8}
\end{figure}
%

%--------------------------------------------------------------
%  Figure 9 (two columns) 
%--------------------------------------------------------------
\begin{figure*}
\centering
\includegraphics[angle=0, width = 16 cm,trim = {50 200 60 140},clip]{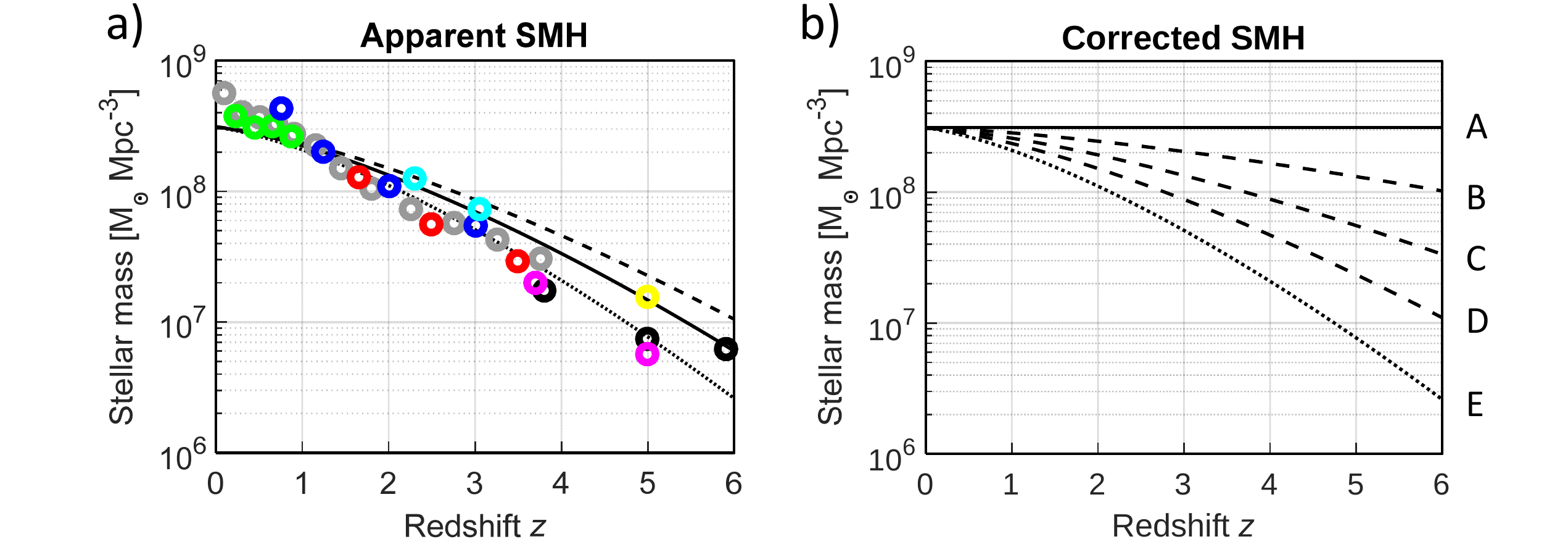}
\caption{
(a) The apparent global stellar mass history (Apparent SMH). The colour circles show observations reported by \citeauthor{Perez_Gonzalez2008}(\citeyear{Perez_Gonzalez2008}, grey), \citeauthor{Pozzetti2010}(\citeyear{Pozzetti2010}, green), \citeauthor{Kajisawa2009}(\citeyear{Kajisawa2009}, blue), \citeauthor{Marchesini2009}(\citeyear{Marchesini2009}, red), \citeauthor{Reddy2012}(\citeyear{Reddy2012}, cyan), \citeauthor{Gonzalez2011}(\citeyear{Gonzalez2011}, black), \citeauthor{Lee2012}(\citeyear{Lee2012}, magenta), and \citeauthor{Yabe2009}(\citeyear{Yabe2009}, yellow). The values are summarized in Table 2 of \citet{Madau2014}. The black lines show the apparent stellar mass history calculated using equation (23) with the UV intergalactic extinction of 0.06 (dashed line), 0.07 (solid line), and $0.085 \,\mathrm{mag}\,h\, \mathrm{Gpc}^{-1}$ (dotted line). The dotted line is also nearly identical with the stellar mass history predicted by a semi-analytic approach of \citet[their Fig. 4, right-hand panel, the solid black line showing prediction for the WMAP5 model]{Somerville2012}. (b) The corrected global stellar mass history (Corrected SMH). The black lines show the stellar mass history after eliminating the effect of the universe opacity assuming $A_{UV}$ of 0.085 (solid line A), 0.065 (dashed line B), 0.045 (dashed line C), 0.025 (dashed line D), and 0.000 $\mathrm{mag}\,h\, \mathrm{Gpc}^{-1}$ (dotted line E).
}
\label{fig:9}
\end{figure*}
%--------------------------------------------------------------
\subsection {Luminosity density evolution}
%--------------------------------------------------------------

The significance of the universe opacity can be tested and verified on modelling of the redshift-dependent luminosity density. Since the evolution of the luminosity density is measured at some frequency $\nu$, we have to modify equation (8) to be frequency dependent:  
\begin{equation}\label{eq17}
I_{\nu} = \frac{c}{H_0} \int_0^{z_{\mathrm{max}}} j_{\nu} \left(z\right)\, \frac{dz}{E\left(z\right)} \, \, ,
\end{equation}
where $I_{\nu}$ is the frequency-dependent EBL, $j_{\nu}\left(z\right)$ is the frequency-dependent luminosity density at redshift $z$ corrected to the dust attenuation
\begin{equation}\label{eq18}
j_{\nu} \left(z\right)\ = j_{\nu 0} \left(1+z\right)^3 e^{-\tau_{\nu}\left(z\right)} \,\, ,
\end{equation}
where $j_{\nu 0}$ is the luminosity density at frequency $\nu$ and at zero redshift, and $\tau_{\nu}\left(z\right)$ is the optical depth at frequency $\nu$ and redshift $z$
\begin{equation}\label{eq19}
\tau_{\nu}\left(z\right) = \frac{c}{H_0} \int_0^{z} \left(\frac{\kappa_{\nu}}{\gamma_0} + \lambda_{\nu 0}\right) \left(1+z'\right)^2 \,\, \frac{dz'}{E\left(z'\right)} \, \, ,
\end{equation}
where $\kappa_{\nu}$ is the galactic opacity at frequency $\nu$, and $\lambda_{\nu 0}$ is the intergalactic attenuation at frequency $\nu$ and at zero redshift. The term $(1+z)^3$ in equation (7) is substituted by $(1+z)^2$ in equation (19) because the opacities depend on the frequency of light at each point of a ray. Since $\kappa_{\nu 0}$ and $\lambda_{\nu 0}$ are rest-frame quantities corresponding to a frequency at the source (but not at the observation point), the opacities decrease along a ray because the wavelengths gradually increase due to the expansion. 

A formula for the optical depth similar to equation (19) has also been reported by \citeauthor{Peebles1993}(\citeyear{Peebles1993}, his Eq. 13.42); \citet{More2009, Johansson2012, Imara2016b} and others. Their derivation is, however, different. It is based on calculating the probability that a light ray intersects dust grains at redshift $z$ in the interval $dz$ when the dust density $n_D$ increases according to equation (7). Optical depth $\tau$ is then obtained by time integration along a ray. This procedure ignores, however, two following effects. First, the arrival rate of photons increases with redshift $z$, so that light absorption by dust per time also increases with $z$. Second, the absorption of photons decreases with $z$ because of its frequency dependence described by the $1/\lambda$ extinction law. Since the two effects are mutually eliminated, the formula of \citet[his Eq. 13.42] {Peebles1993}; \citet[their Eq. 12]{More2009} and other authors is identical with equation (19).

Figs.~\ref{fig:7} and ~\ref{fig:8} show the optical depth, extinction correction, and observed and predicted UV luminosity densities as a function of redshift for $z < 10$. The intergalactic attenuation at UV wavelengths is in the range of $0.06 - 0.085 \,\mathrm{mag} \, h\, \mathrm{Gpc}^{-1}$ being about three times higher than attenuation at visual wavelengths  $\lambda_V = 0.02 \,h\, \mathrm{Gpc}^{-1}$ \citep{Xie2015}. The optical depth increases from 0 to 10 (Fig.~\ref{fig:7}a,b), while the extinction correction (i.e. the coefficient correcting the luminosity) drops from 1 to $1\times10^{-5}$ (Fig.~\ref{fig:7}c,d) in the range of redshifts 0 < z < 10. Considering this attenuation in equation (18), we reproduce a steep increase of the luminosity density at low redshifts, the position of the maximum at $z \sim 3$ as well as the decrease of the luminosity density at high redshifts (Fig.~\ref{fig:8}). 

The increase of the luminosity density at low $z$ in Fig.~\ref{fig:8} is produced by a transformation from the comoving to the proper volume of the Universe in the luminosity integral (17). Physically, the increase originates in the fact that the Universe occupied a smaller volume in previous epochs. The decline of luminosity at redshifts $z > 3$ in Fig.~\ref{fig:8} is caused by intergalactic attenuation which rapidly increases with redshift and causes the Universe to be significantly opaque. This result questions the standard interpretation of the luminosity density evolution as a direct consequence of the evolution of the star formation rate in the Universe. As shown above, the relation between the luminosity density evolution and the star formation rate is not as simple as assumed and the culmination of the luminosity density at $z \sim 3$ does not necessarily mean a high star formation rate at this epoch.

%--------------------------------------------------------------
\subsection {Global stellar mass density history}
%--------------------------------------------------------------

So far, we have assumed a constant comoving galaxy number density with cosmic time. If the comoving galaxy number density evolves with redshift, the luminosity density $j_{\nu 0}=j_{\nu 0}\left(z\right)$ is expressed using a redshift-dependent global stellar mass density $\rho \left(z\right)$ and the luminosity of the Sun $L_{\Sun}$
\begin{equation}\label{eq20}
j_{\nu 0}\left(z\right) = \rho_0\left(z\right) L_{\Sun} \, \, ,
\end{equation}
and equation (18) reads
\begin{equation}\label{eq21}
j_{\nu}\left(z\right) = \rho\left(z\right)(1+z)^3 L_{\Sun}\, e^{-\tau_{\nu}{\left(z\right)}} \,\, .
\end{equation}
This equation can be used for determining the stellar mass density history $\rho\left(z\right)$ using observations of the luminosity density $j_{\nu}\left(z\right)$ and optical depth $\tau_{\nu}\left(z\right)$.

Since we consider the global stellar mass-to-light ratio constant with cosmic time in equation (20), the obtained $\rho\left(z\right)$ might not be very accurate at high redshifts \citep{Papovich2001, Conroy2013, Madau2014}. However, the errors due to this simplification are probably lower than those produced by uncertainties in current observations of the luminosity density $j_{\nu}\left(z\right)$ and optical depth $\tau_{\nu}\left(z\right)$.
	
The star formation rate and the global stellar mass density history have been determined and interpreted from observations of the luminosity density evolution by many authors, see e.g., \citet{Lilly1996, Madau1998, Dickinson2003, Hopkins2006, Somerville2008, Somerville2012, Madau2014}. Their approach is, however, different because the attenuation term in equation (21) is neglected 
\begin{equation}\label{eq22}
j_{\nu}\left(z\right) = \rho^A\left(z\right)(1+z)^3 L_{\Sun}  \, \, .
\end{equation}
In this way, they obtain an 'apparent' stellar mass density  $\rho^A\left(z\right)$, which is biased because it includes the redshift-dependent universe opacity 
\begin{equation}\label{eq23}
\rho^A\left(z\right) = \rho\left(z\right) e^{-\tau_{\nu}{\left(z\right)}} \,\, .
\end{equation}

Fig.~\ref{fig:9}a shows observations of the apparent stellar mass density $\rho^A\left(z\right)$ (colour circles) together with theoretical predictions calculated by equation (23) for a constant $\rho$ and three alternative levels of intergalactic attenuation at UV wavelengths in the range of $0.06 - 0.085\,  \mathrm{mag} \,h\,\mathrm{ Gpc}^{-1}$ used for fitting the luminosity density evolution in Fig.~\ref{fig:8}. Fig.~\ref{fig:9}a demonstrates that the exponential decay of the apparent stellar mass density $\rho^A\left(z\right)$ can fully originate in the universe opacity. If the true universe opacity is weaker than that assumed in the modelling, the true stellar mass density must exponentially decline with redshift. The actual value of this decline is obtained after correcting $\rho^A\left(z\right)$ for attenuation (see Fig.~\ref{fig:9}b for several possible scenarios).

%--------------------------------------------------------------
\section {Discussion}
%--------------------------------------------------------------

The theoretical analysis of the EBL disproves the opinion that the Olbers' paradox is eliminated by considering a model of expanding universe of finite age \citep{Wesson1987,Wesson1989,Wesson1991}. The calculations show that the finite age of the Universe is not a decisive factor. The EBL diverges provided the Universe and galaxies are transparent, and the number density of galaxies in the comoving volume and the mean galaxy luminosity do not change with cosmic time. The EBL divergence is produced by radiation of galaxies at high redshifts when the galaxies are concentrated in a small volume. The divergence is removed by the universe opacity and/or by the universe darkness at high redshifts.  

The opacity of galaxies and intergalactic space is caused by light absorption by dust. Dust in galaxies affects mostly the stellar EBL. It causes a partial opacity of foreground galaxies and thus reduces the intensity of stellar light from distant background galaxies. The stellar energy absorbed by dust is further re-emitted at the IR and FIR wavelengths. Hence, the galactic dust affects the spectral characteristics of the EBL by transforming the stellar to IR and FIR light but has little effect on the total EBL. By contrast, the impact of the intergalactic dust on the EBL is different. The intergalactic dust causes non-zero opacity of the Universe which is quite minute being about 0.01 mag  $h \,  \mathrm{ Gpc}^{-1}$, so the local Universe appears effectively transparent. However, the opacity increases with redshift and the Universe becomes considerably opaque at redshifts $z > 3$. This increase has been confirmed by observations and supported by theoretical works of several authors \citep{Menard2010a, Johansson2012, Xie2015, Imara2016b}. Since the intergalactic dust is colder than the galactic dust, the absorbed EBL is re-radiated in the micro-wave spectrum. 

The calculations of the EBL in the Universe with redshift-dependent opacity yield a satisfactory fit with observations. The predicted bolometric EBL of $100-200 \,\mathrm{nWm}^{-2}\mathrm{sr}^{-1}$ is within the limits of the observed values of $40-200 \,\mathrm{nWm}^{-2}\mathrm{sr}^{-1}$ in the band of wavelengths from 0.1 to 1000 $\mu$m. Similarly, the predicted luminosity density evolution fits the luminosity measurements (Fig.~\ref{fig:8}). The model reproduces a steep increase of the luminosity density at $z < 2$, its maximum at $z = 2-3$, and finally its decrease at higher redshifts. The increase of the luminosity density at low $z$ does not originate in the evolution of the star formation rate as commonly assumed but in the change of the proper volume of the Universe which must be considered in the luminosity integrals. The decrease of the luminosity density at high $z$ originates in the opacity of the early Universe. 

By contrast, current calculations of the EBL neglect the universe opacity and assume a transparent universe even at high redshifts. The observed luminosity density evolution is ascribed to the evolution of the global stellar mass density in the Universe \citep{Madau1996, Madau1998, Hopkins2004, Hopkins2006, Bouwens2007, Bouwens2011, Bouwens2015, Dominguez2011, Gilmore2012, Somerville2012}. The EBL divergence is eliminated by darkness of the early Universe caused by decline of the global stellar mass density at high redshifts. 

To figure out whether and to which extent the early Universe was opaque and/or dark is intricate because of trade-off between the both phenomena. The observations of the evolution of the galaxy number density and the global stellar mass density are not decisive. They are based on measurements of the overall effect of the universe opacity and darkness with no power to separate them. The observations of the universe opacity are more convincing \citep{Menard2010a, Johansson2012,Xie2015, Imara2016b} but still display large uncertainties. Considering the opacity within the current error limits, the acceptable models predict a much lower decline of the global stellar mass density with redshift than commonly assumed. The decline is even negligible for the universe opacity of $A_{UV} = 7 - 8.5 \times 10^{-2} \, \mathrm{mag} \, h \, \mathrm{Gpc}^{-1}$. If the true universe opacity is lower than this value, the early Universe ($2 < z < 10$) is partially dark and partially opaque. In this case, we have to figure out why the decline of the global stellar mass density with redshift obeys the same exponential law as for attenuation.

For resolving the balance between the universe opacity and darkness, we need more accurate measurements of the universe opacity and further detailed studies of the luminosity, number density and stellar mass evolution of high-redshift galaxies. For example, observations of a different growth of the number and stellar mass densities of massive quiescent and star-forming galaxies from high to low redshift reported by \citet{Hopkins2010, vanDokkum2010, Brammer2011, Tomczak2014} and others, might have partly origin in the frequency- and redshift-dependent universe opacity and can provide some constraints on it. Obviously, interpretations of the luminosity density evolution in terms of the universe opacity could revise our understanding of the global star formation and stellar mass histories at the early epoch of the Universe.

In order to keep the problem simple, we focused on effects with a major impact on the EBL. For example, we ignored gravitational lensing in the EBL calculations. The lensing can affect radiation by focusing or defocusing beams of light and produces local perturbations of the EBL \citep{Bartelmann2001}. However, since the average flux must be conserved \citep{Peacock1986, Peacock1999}, the overall value of the EBL should be unaffected.  We also avoided analysing the spectral distribution of the EBL and studied the UV luminosity density evolution and the bolometric EBL only. Since the bolometric luminosity of galaxies is probably less sensitive to their age, we partly eliminated the problems related to the evolution of galaxies and to evolution of their spectral energy distributions. Obviously, predicting the spectral energy distribution of the EBL is a more complicated task in which the evolution of galaxies in time cannot be ignored \citep{Dominguez2011, Primack2011, Gilmore2012}. Hence, a more accurate modelling of the EBL should include calculations of the spectral energy distribution of the EBL considering wavelength-dependent galactic and universe opacities, evolution of galaxies and knowledge of the history of the universe expansion including its acceleration.

\section*{Acknowledgements}
I thank an anonymous reviewer for helpful comments and Alberto Dom\'{i}nguez for providing me kindly with Fig.~\ref{fig:3}.

%%%%%%%%%%%%%%%%%%%%%%%%%%%%%%%%%%%%%%%%%%%%%%%%%%

%%%%%%%%%%%%%%%%%%%% REFERENCES %%%%%%%%%%%%%%%%%%

% The best way to enter references is to use BibTeX:

\bibliographystyle{mnras}
\bibliography{paper} % if your bibtex file is called example.bib

%%%%%%%%%%%%%%%%%%%%%%%%%%%%%%%%%%%%%%%%%%%%%%%%%%

% Don't change these lines
\bsp	% typesetting comment
\label{lastpage}
\end{document}